\documentclass[12pt]{iopart}

\usepackage{subfigure}

\usepackage[dvips]{graphicx,color}
\usepackage{dcolumn}
\usepackage{bm}
\usepackage{ulem}
\usepackage{soul}

\begin{document}

\title{Nanostructures made from superconducting boron doped diamond}

\author{Soumen Mandal\dag\ \S\ , C\'ecile Naud\dag\ , Oliver A.
Williams\ddag\ , \'{E}tienne Bustarret\dag\ , Franck
Omn\`{e}s\dag\ , Pierre Rodi\`{e}re\dag\ , Tristan Meunier\dag\ ,
Laurent Saminadayar\dag\ \P\ \S\ and Christopher B\"auerle\dag\
\footnote[3]{Address of Correspondence:
bsm\_all@listes.grenoble.cnrs.fr}}

\address{\dag\ Institut N\'eel, CNRS and Universit\'{e} Joseph Fourier,
38042 Grenoble, France}%
\address{\ddag\ Fraunhofer Institut Angewandte
Festk\"{o}rperphysik, Tullastra{\ss}e 72, 79108 Freiburg, Germany}
\address{\P\ Institut Universitaire de France, 103 boulevard
Saint-Michel, 75005 Paris, France} %

\begin{abstract}
We report on the transport properties of nanostructures made from
boron-doped superconducting diamond. Starting from nanocrystalline
superconducting boron-doped diamond thin films, grown by Chemical
Vapor Deposition, we pattern by electron-beam lithography devices
with dimensions in the nanometer range. We show that even for such
small devices, the superconducting properties of the material are
well preserved: for wires of width less than $100\, nm$, we
measure critical temperatures in the Kelvin range and critical
field in the Tesla range.
\end{abstract}


\maketitle

\section{Introduction}
The discovery of superconductivity\cite{Nagamatsu} in $MgB_2$ has
generated a lot of interest for a special class of superconducting
materials belonging to the covalent metals. In this context, the
observation of superconductivity in highly-doped boron
diamond\cite{Ekimov} paved the way to the study of superhard
superconducting materials\cite{Blase_09,Dubitskiy}. Apart from the
fundamental point of understanding of the physical mechanisms
leading to the superconductivity in these systems, their interest
lies in a very high Young's modulus, which makes them promising
candidates for the fabrication of superconducting Nano
Electro-Mechanical Systems of exceptional quality factor.
Evidences that \textsl{both} the
superconductivity\cite{Nesladek,Gajewski,Achatz} \textsl{and} the
mechanical properties\cite{Imboden} are essentially preserved in
nanocrystalline layers grown on non-diamond substrates such as
silicon have to be checked in order to bring further credit to
this approach.

Most of the studies, however, have so far focused on the
properties of \textsl{bulk} superconducting diamond, and it still
remains to be shown what happens when the material is patterned
into nanostructures. In this paper, we present a comprehensive
study of nanostructured superconducting polycrystalline diamond
films. Our measurements show that the critical temperature of the
nanostructures being $\approx$2.5 K while for the bulk this is
$\approx$3.5 K. The critical field for these structures is
approximately $500$mT at 50mK.

\section{Experimental}
The nanocrystalline boron-doped diamond was obtained by Chemical
Vapour Deposition on a silicon oxide wafer. Prior to growth,
wafers were cleaned with $NH_{3}OH : H_{2}O_{2} : H_{2}O$
($1:1:5$) solution at $75^{\circ}C$ for $10\, min$, and rinsed in
pure DI water in an ultrasonic bath. In order to enhance
nucleation, wafers were then seeded with diamond nano-particles
from an aqueous colloid of mono-disperse diamond particles known
to have sizes less than $7\, nm$ in solution as confirmed by
dynamic light scattering\cite{Williams,Villar}. Atomic Force
Microscope (\textsc{afm}) measurements have shown this technique
to result in uniform nucleation densities far in excess of
$10^{11}\, cm^{-2}$. Diamond growth was then performed by
microwave plasma enhanced chemical vapour deposition
(\textsc{mpcvd}) from $4\%$ $CH_4$ diluted in $H_2$ with
additional boron from a trimethylboron gas source. The microwave
power was $3000\, W$ at $60\, mbar$, the substrate temperature
being around $800^{\circ}C$ as monitored in situ with a
bichromatic pyrometer.

\begin{figure}\begin{center}\includegraphics[width=8cm,angle=0]{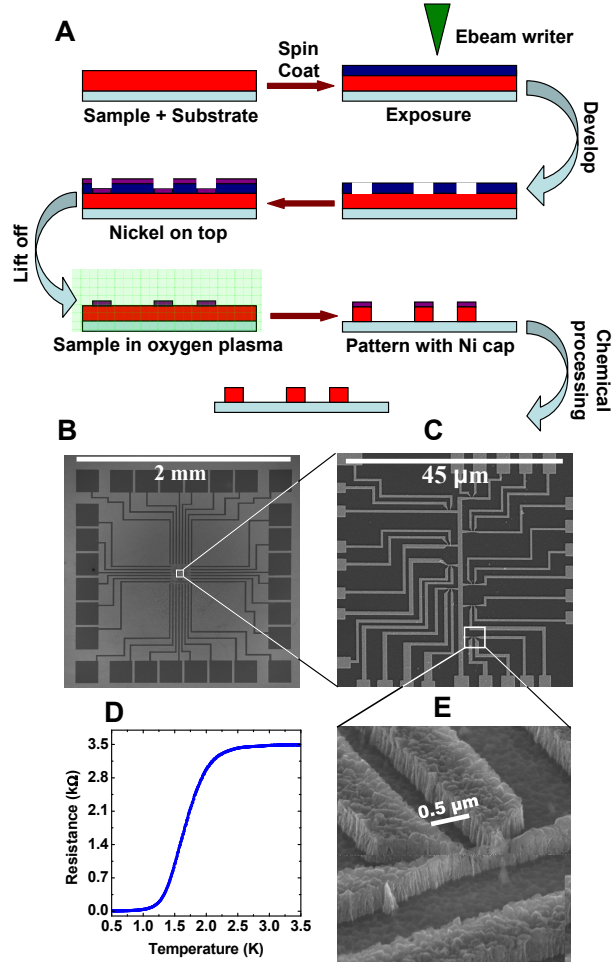}\end{center}
\caption{Schematics of the nanofabrication process with
\textsc{sem} images of a typical microcircuit. Panel B shows the
\textsc{sem} micrograph of the microcircuit used in our
experiments. Panel C shows the magnified view of a ${50}\times{50}
{\mu m}^2$ area in the middle of the microcircuit. Panel E is the
slanted view of the thinnest line we obtained which is $500\, nm$
long, $90\, nm$ wide and $300\, nm$ high. The $R$ vs $T$ curve for
this wire is shown in panel D: a clear superconducting transition
is observed at $T \approx 1.7\, K$.} \label{scheme}
\end{figure}

Nanostructured devices have then been fabricated from this
material. In figure \ref{scheme} we present a diagram of the
entire nanofabrication process. The wafer was first spin-coated
with $4\,\%$ polymethyl-methacrylate (\textsc{pmma}) to form a
$250\, nm$ thick layer and prebaked at $180^{\circ}C$ for $5\,
min$. The next step was the exposure to the electron beam with a
dose of $360\,\mu C\cdot cm^{-2}$ for an acceleration voltage of
$20\, kV$. After exposure, the pattern was developed in $1:3$
solution of methyl isobutyl ketone (\textsc{mibk}) and iso-propyl
alcohol (\textsc{ipa}) for $1\, min$. A thin layer of nickel
($65\, nm$) has then been deposited and patterned using a standard
electron-gun evaporator and lift-off technique. This $Ni$ layer
acts as a mask for the plasma etching of the diamond structures
subsequently done using Electron Cyclotron Resonance oxygen
plasma\cite{Bernard} and a $-27\, V$ $dc$ bias for $\approx 8\,
min$. This leads to an etching rate of $\approx 40\,nm/min$. This
mask was finally removed using an $FeCl_3$ solution. Ohmic
contacts were then obtained by an evaporation of metals ($Ti-Au$).

Panel B, C and E of figure \ref{scheme} show \textsc{sem} pictures
of the obtained nanostructures, namely wires of different widths
and lengths. Panel E represents the smallest device we have been
able to fabricate, a wire of length $500\, nm$ and width $90\,
nm$; note that in this case the aspect ratio is as high as
$\sim1:3$, the anisotropy of the plasma etching allowing to
pattern one single grain. The superconducting transition of this
wire is presented in panel D showing a critical temperature of
$\approx 1.7$K. Panel B in figure \ref{scheme} shows a typical
microcircuit used for our measurements and panel C is the blow-up
of a ${50}\times{50}\, {{\mu}m}^2$ area in the center of the
microcircuit.

\section{Results and Discussion}
\begin{figure}\begin{center}\includegraphics[width=8cm,angle=0]{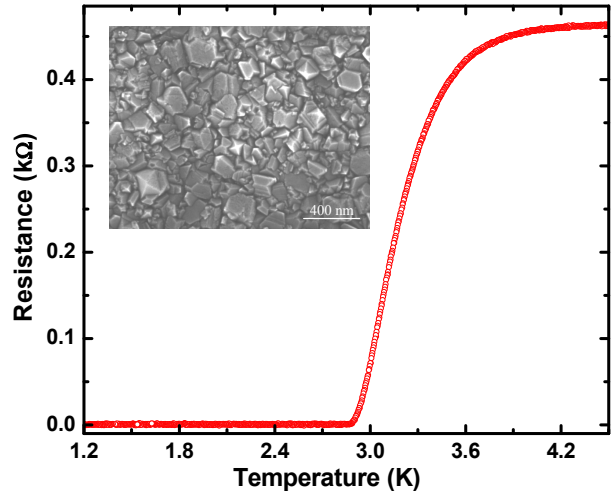}\end{center}
\caption{Superconducting resistive transition in a boron-doped
diamond thin film. The inset is a Scanning Electron Microscope
(\textsc{sem}) image of the surface of the sample, consisting of
grains of typically $\approx 150\, nm$.} \label{bulk}
\end{figure}

A Scanning Electron Microscope (\textsc{\textsc{sem}}) picture
(inset of figure \ref{bulk}) of the sample shows grains of typical
size $\approx 150\, nm$ for a film thickness of $\approx 250\,
nm$, consistent with the nucleation density. Four silver paste
electrical contacts, about $5\, mm$ apart from each other, have
been deposited at the surface for the characterisation of the
as-grown layer. The 4-points resistance was measured as a function
of temperature using these contacts and a standard $ac$ lock-in
technique under a very low current injection of $1\mu A$. Typical
data are presented in figure \ref{bulk}: a clear superconducting
transition is observed with a zero resistance at $\approx 3\, K$.
The width of the transition is quite large, typically $0.7\, K$
with a $10\,\% - 90\,\%$ of the onset resistance criterion; we
attribute this width to the distribution of the sizes of the
grains in the material.

\begin{figure}\begin{center}\includegraphics[width=7cm,angle=0]{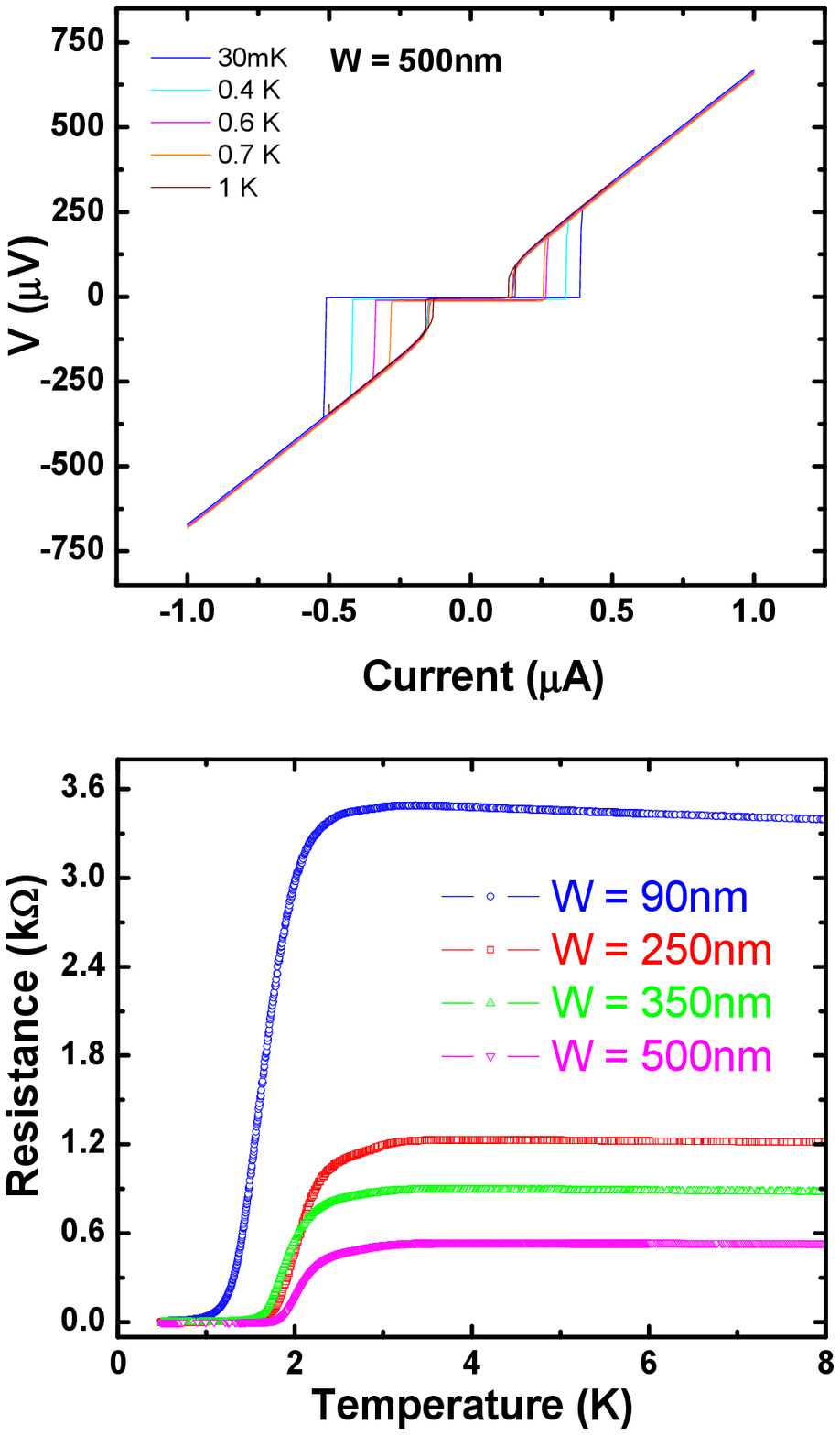}\end{center}
\caption{Voltage-current ($V-I$) characteristic of a $500\, nm$
wide wire at different temperatures are shown in top panel. The
$V-I$ curves are hysteretic due to thermal effect (Joule heating).
The bottom panel shows the $R$ vs $T$ curves for four
representative wires.} \label{ivt}
\end{figure}

\begin{figure}\begin{center}\includegraphics[width=7cm,angle=0]{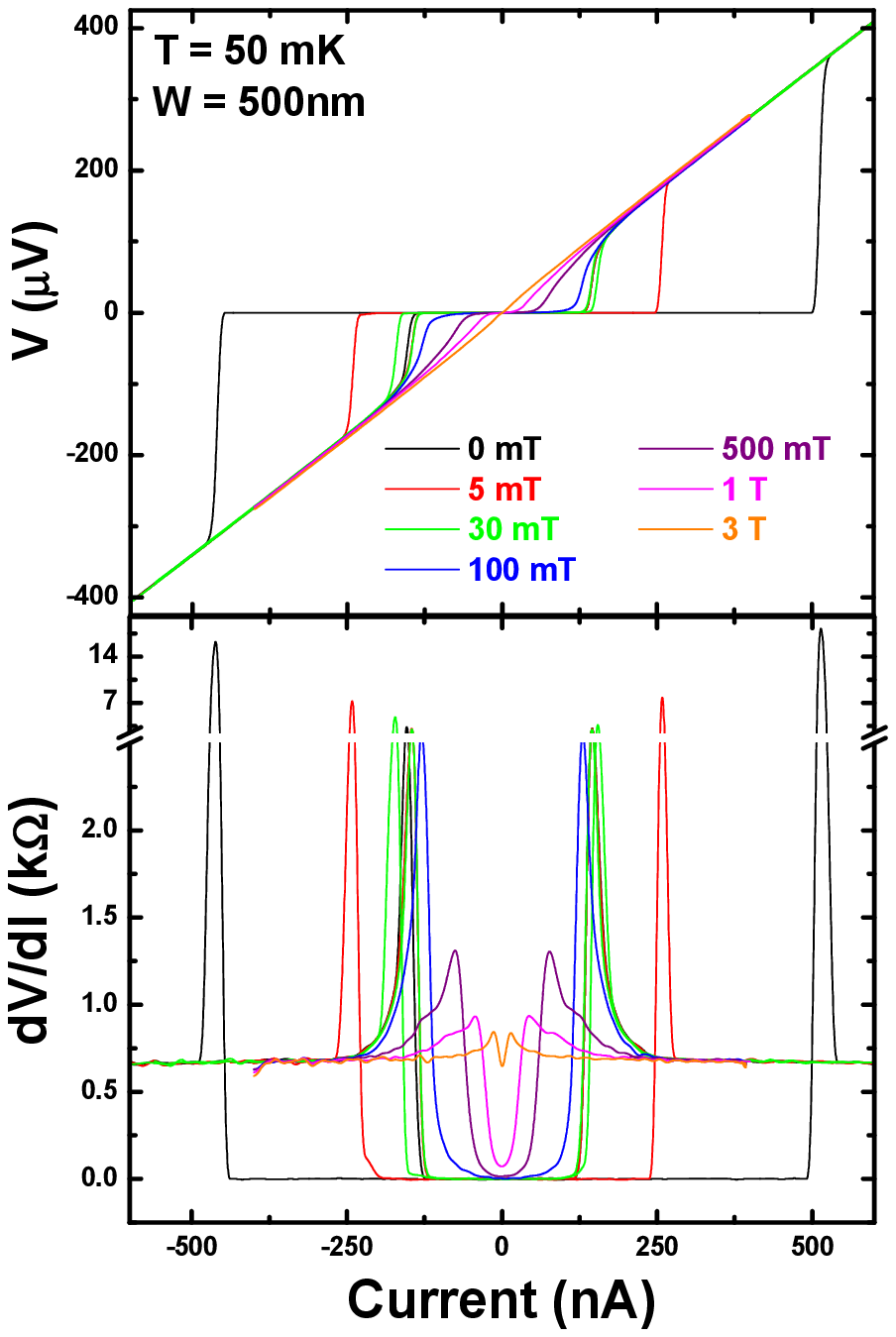}\end{center}
\caption{Voltage-current ($V-I$) characteristic of a $500\, nm$
wide  wire at $50\, mK$ under different magnetic fields. The field
is applied perpendicular to the plane of the sample. Only
the~\textquotedblleft\textsl{increasing}\textquotedblright~parts
of the characteristics have been plotted. The behaviour is
hysteretic until the applied field reaches $30\, mT$.  The bottom
panel shows the differential resistance extracted from the $V-I$
curves. The resistance goes to zero when the wire is in its
superconducting state.} \label{ivb}
\end{figure}

Electrical characterisation of the devices were performed in both
a  ${^{3}He}$ and a ${^{3}He} / {^{4}He}$ dilution refrigerators.
Bottom panel of figure \ref{ivt} shows the critical temperature
$T_c$ of wires of various widths, measured with very low current
(typically $100\, nA$); no significant difference was observed
from the critical temperature measured on the~\textquotedblleft
bulk\textquotedblright~sample ($\approx 2\, K$ for this wafer)
except for the case of the narrowest wire (below $100\, nm$ wide,
$T_c \approx 1.7\,K$). This is a generic observation for all our
samples: the critical temperature of our wires is those of the
bulk material except for wires thinner than typically $\approx 100
\, nm$; in this case, $T_c$ it is slightly reduced. Figure
\ref{ivt} shows the voltage-current ($V-I$) characteristic of a
$500\, nm$ wide wire measured at different
temperatures\cite{ivreason}. The $V-I$ curves are hysteretic due
to thermal effect: when the critical current is reached, Joule
effect heats up the wire and the critical current measured when
subsequently \textsl{decreasing} the current is thus much lower
than the critical current measured when \textsl{increasing} the
current\cite{Rabaud,Courtois}; moreover, this~\textquotedblleft
retrapping\textquotedblright~current is independent of the
temperature of the refrigerator, as the \textsl{actual}
temperature of the sample is then fixed by the $dc$ current trough
the wire.

Critical field measurements were performed by applying a magnetic
field perpendicular to the structure. $V-I$ characteristics for a
$500\, nm$ wide wire under different magnetic fields and at $50\,
mK$ are presented on the top panel of figure \ref{ivb}. As
expected, the critical current decreases when applying a magnetic
field. It should be stressed that for magnetic fields larger than
$30\, mT$, the $V-I$ characteristic is not hysteretic: in this
case, the Joule heating becomes negligible as the critical current
is strongly lowered, and both
(\textquotedblleft\textsl{increasing}\textquotedblright~and~\textquotedblleft\textsl{decreasing}\textquotedblright)
critical currents become similar. In order to get more insights on
the superconductivity of our devices under magnetic field, we have
numerically calculated the derivatives of the $V-I$
characteristics: the obtained ${\frac{dV}{dI}}(I)$ curves are
displayed on the bottom panel of figure \ref{ivb}. For magnetic
fields smaller than $500\, mT$, a clear zero is observed around $I
= 0$, as expected for a superconductor. More interestingly, we
observe a dip at zero current in the ${\frac{dV}{dI}}(I)$
characteristic for magnetic fields above $3\, T$. This
demonstrates that one can still observe traces of
superconductivity above 3T in our nanometric wires, the most
probable mechanism being that some grains remain superconducting
under very high magnetic fields. More local studies, such as
Scanning Tunneling Spectroscopy measurements\cite{Sacepe,Dahlem},
are certainly desirable to fully understand the robustness of the
superconductivity observed in our system.

\section{Conclusion}
In conclusion, we have successfully fabricated nanostructures from
boron-doped nanocrystalline superconducting diamond. Using
electron beam lithography, we have fabricated devices of
characteristic size less that $100\, nm$ and aspect ratio as high
as $1:3$. These structures have critical temperatures in the
Kelvin range, similar to what is observed in ``bulk"  films;
typical bulk T$_c$ being 3.5K while for nanostructures it is 2.5K
except for 90nm wire where it is $\approx$1K. We measure critical
fields close to $500\, mT$ and traces of superconductivity are
observed above $3\, T$. This study proves that superconductivity
in boron-doped diamond is a very robust phenomenon which makes it
a promising candidate for future applications in the field of
superconducting Nano Electro-Mechanical Systems.

\ack We thank L. Marty for help in the \textsc{afm} measurements
and P. Mohanty and M. Imboden for fruitful discussions. This work
has been supported by the French National Agency (\textsc{anr}) in
the frame of its programme in~\textquotedblleft Nanosciences and
Nanotechnologies\textquotedblright~(\textsc{supernems} project
$n^{\circ}\textsc{anr}-08-\textsc{nano}-0 33$). O.A.W. acknowledges financial support from the Fraunhofer Attract award ``Hybrid HF-MEMS Filters for GHz-Communication and capillary MEMS systems for chemical and bio-chemical Sensing - COMBIO".

\section*{Refrences}


\begin{thebibliography}{99}
\bibitem{ivreason}
The asymmetry seen in the VI characteristics mainly arises due to
an offset in our experimental setup, however, there is also a
residual asymmetry which is presently not understood.
\end{thebibliography}
\end{document}